\documentclass[prd,twocolumn,showpacs,floatfix]{revtex4}
\usepackage{bm}
\usepackage{amssymb}
\begin{document}
\title{Survival of the black hole's Cauchy horizon under non-compact
perturbations}
\author{Lior M.\ Burko}
\affiliation{Department of Physics, University of Utah, Salt Lake City,
Utah 84112}
\date{June 4, 2002}
\begin{abstract}
We study numerically the evolution of spactime, and in particular of a
spacetime singularity, inside a black hole under a class of perturbations
of non-compact support. We use a very simplified toy model of a spherical
charged black hole which is perturbed nonlinearly by a self-gravitating, 
spherical scalar field. The latter grows logarithmically with
advanced time along an outgoing characteristic hypersurface. We find that
for that class of perturbations a portion of the Cauchy horizon survives
as a non-central, null singularity.
\end{abstract}
\maketitle

{\it Introduction and summary}: The geometrical and physical properties of
the Cauchy horizon singularity inside black holes have received much
attention \cite{review}. That singularity was shown to be null,
non-central, and weak. The weak nature of the Cauchy horizon singularity
has far-reaching implications. In particular, it leaves open the
possibility that physical objects which fall into a black hole may
traverse the Cauchy horizon singularity only mildly affected, and
re-emerge in another universe. 

The evolution of spacetime geometry into a (weak) curvature singularity at
the Cauchy horizon has been studied both numerically and analytically 
\cite{poisson-israel90,ori91,ori92,brady-smith95,burko97,burko-ori98,
burko99a,remark}. In all these studies the black hole was taken to be
isolated, and the source of the perturbations was taken to be the
perturbations which result from the evolution of nonvanishing multipole
moments during the collapse. These perturbations are inherent to any
nonspherical gravitational collapse, and result from the backscattering of
waves, which are created during the collapse, off the curvature of
spacetime \cite{price}. These perturbations have a compact support at some
initial time.

It is interesting to ask whether the evolution of a null and weak
singularity at the Cauchy horizon is just an artifact of the assumption of
compactness. That is, will any dominating perturbation field which has
non-compact
support on the initial time slice lead to the full destruction of the null
singularity, and its replacement by a spacelike one? In fact, the
perturbations due to the collapse and the resulting tails can be thought
of as a lower bound on the perturbation field. It is intersting to
ask what happens to the Cauchy horizon if perturbations which are stronger
than that lower bound are present. 

This question is interesting not just from the mathematical
viewpoint: indeed, a generic class of perturbations exists, where the
perturbation field has non-compact support. These are the perturbations
which arise from the capture of photons which originate from the relic
cosmic background radiation (CBR). Even if removed from any conceivable
astrophysical object, any black hole is still perturbed by the CBR. Unlike
the perturbations due to the collapse, the CBR perturbations are
non-compact. Because the perturbation field is greater than the lower
bound set by the perturbations due to the tails, one may suspect that the
evolution of spacetime, and in particular of the singularity, be
dominated by the non-compact perturbations, rather than by the compact
ones. 

If this is indeed the case, the non-compact perturbations threaten
to change our notions of the causal structure inside black holes. When the
perturbations due to the tails are considered, it is found that the
weakness of the Cauchy horizon singularity is a rather delicate issue: it
depends on certain integrals being bounded. Specifically, assume that
the field is due to the tails. Then, on
the event horizon the scalar field $\Phi$ behaves like $\Phi=(\kappa
v)^{-n}$, where $\kappa$ is a constant, $v$ is advanced time,
and $n$ is a positive integer which is related to the multipole
moment of the perturbation field. Denoting
schematically by ${\cal R}$ the fastest growing components of the
Riemann-Christoffel tensor approaching the Cauchy horizon, the curvature
at that limit behaves like ${\cal R}\approx
\tau^{-2}[-\ln(-\tau)]^{2n+2}$, where $\tau$ is proper time along a
timelike geodesic which is set equal to zero on the Cauchy horizon. The
Cauchy
horizon singularity is weak, if ${\cal R}$ is twice-integrable. (This last
statement can be made precise \cite{clarke-krolak}.) For positive value of
$n$ this is indeed the case. However, small changes in ${\cal R}$, e.g., 
another factor of $\tau^{-\epsilon}$ for any small and positive $\epsilon$,
would change the picture entirely, as ${\cal R}$ would no longer be twice
integrable. That is, the twice integrability of ${\cal R}$ is strongly
dependent on the form of the field at the event horizon. The twice
integrability of ${\cal R}$ (and consequently the weakness of the Cauchy
horizon singularity) depends then on the assumption that the scalar field
has a compact support (as this condition leads, through Price's analysis
\cite{price}, to the tail form for the field on the event horizon). 
Dominating non-compact perturbations threaten to change ${\cal R}$ in a
significant way, such that it would no longer be twice integrable. 
One can ask then the following question: Is the Cauchy horizon necessarily
utterly destroyed and replaced by a spacelike singularity when
perturbations with non-compact support are present, or can it still
survive (as a null, weakly singular hypersurface) also when perturbations
of non-compact support are present? 

In this paper we shall answer the
latter question in the affirmative. We show that a certain class of
non-compact perturbations still preserves the null, non-central nature of
the Cauchy horizon singularity, even though the evolution of geometry and
of the singularity is indeed dominated by the non-compact
perturbations. We emphasize that any perturbation field with non-compact
support (whose dynamics dominates deep inside the black hole) is
appropriate, as it serves as a counter-example for the claim that only
perturbations with compact support can evolve into a null and weak
singularity. The survival of the Cauchy horizon as a null, non-central
singularity can occur also when perturbations of non-compact support
(of certain classes) are present. 

{\it Model}: In this paper we study the evolution of spacetime curvature
inside a black hole in the presence of perturbations which have
non-compact support under a very simplified toy model. For simplicity, we
take the black hole to be spherically symmetric, and to have a fixed
electric charge $Q$. This is a useful toy model for a spinning black hole,
because the unperturbed spacetimes, namely the Reissner-Nordstr\"{o}m and
Kerr spacetimes, respectively, have very similar causal structures, which
lead to similar blue-sheet effects near their inner horizons. In
fact, much of the understanding we currently have about black hole
interiors have been obtained through the study of spherical charged
models. (One important difference is that the null singularity inside a
spherical charged black hole is monotonic, whereas the one inside a
spinning black hole is oscillatory \cite{ori99}. This difference is not
crucial for our purposes here. Another difference is related to the
question of the occurrence of a spacelike singularity inside black
holes. A spacelike singularity to the future of the Cauchy horizon 
singularity was found in spherically-symmetric, charged models. It has
been argued that no corresponding spacelike singularity is likely to occur
inside a rotating black hole \cite{ori99}. Others have argued, that a
spacelike singularity, possibly of the Belinskii-Khalatnikov-Lifshitz
type, is a possible outcome. While this open question is extremely
important, it is unrelated to the nature of the null singularity which
precedes the spacelike one, if such a spacelike singularity exists.)

We write the spherically-symmetric metric in double-null
coordinates in the form
\begin{equation}
\,ds^2=-2e^{2\sigma (u,v)}\,du\,dv+r^2(u,v)\,d\Omega^2 \,
\label{metric}
\end{equation}
where $\,d\Omega^2$ is the line element on the unit two-sphere. 
As the source term for the Einstein equations, we take the contributions
of both the scalar field $\Phi$ and the (sourceless) spherical electric
field (see \cite{burko-ori97} for details). The dynamical equations are
the scalar field equation $\nabla_{\mu}\nabla^{\mu}\Phi=0$ and the
Einstein equations, which
reduce to  
\begin{eqnarray}
\Phi_{,uv}+\frac{1}{r}\left(r_{,u}\Phi_{,v}+r_{,v}\Phi_{,u}\right)=0
\label{KGEQ}
\end{eqnarray}
\begin{eqnarray}
r_{,uv}+\frac{r_{,u}r_{,v}}{r}+\frac{e^{2\sigma}}{2r}\left(1-\frac{Q^{2}}
{r^{2}}\right)=0
\label{EEQ1}
\end{eqnarray}
and
\begin{eqnarray}
\sigma_{,uv}-\frac{r_{,u}r_{,v}}{r^2}-\frac{e^{2\sigma}}{2r^2}
\left(1-2\frac{Q^{2}}{r^{2}}\right)+
\Phi_{,u}\Phi_{,v}=0 \; .
\label{EEQ2}
\end{eqnarray}   
These equations are 
supplemented by the two constraint equations
\begin{eqnarray}
r_{,uu}-2\sigma_{,u}r_{,u}+r(\Phi_{,u})^{2}=0
\label{con1}
\end{eqnarray}
\begin{eqnarray}
r_{,vv}-2\sigma_{,v}r_{,v}+r(\Phi_{,v})^{2}=0.
\label{con2}
\end{eqnarray}
Although similar, there is an important difference between the numerical
evolution of a code based on Eqs.~(\ref{KGEQ})-(\ref{EEQ2}) and a code
which is based on the dynamical equations used in
Ref.~\cite{burko-ori97}: In the latter case, the wave equation
for the $g_{uv}$ metric function becomes a free wave equation
asymptotically close to the Cauchy horizon. (Notice that $g_{uv}$ vanishes
exponentially in $v$ near the Cauchy horizon, whereas
$r_{,u},r_{,v},\Phi_{,u}$ and $\Phi_{,v}$ decay like inverse powers of
$v$, where $v$ is proportional to advanced time -- see below.) This
implies that the numerical
integration becomes inaccurate near the Cauchy horizon because
dynamically-important terms become negligible. When the field equations
are written as Eqs.~(\ref{KGEQ})-(\ref{EEQ2}) this problem does not occur.

{\it Initial value problem}: From the pure initial-value viewpoint, we
need to specify three initial functions on each segment of the initial
surface: $r$, $\sigma$, and $\Phi$. The constraint equations reduce this
number: Eqs.~(\ref{con1}) and (\ref{con2}) impose one constraint each on
the initial data at $u=u_i$ and $v=v_i$, respectively. The remaining two
initial functions, however, represent only one physical degree of
freedom: The other degree of freedom expresses nothing but the gauge
freedom associated with the arbitrary coordinate transformation $u\to
\tilde u(u)\;,\;v\to \tilde v(v)$. In what follows we shall use a standard
gauge, in which $r$ is linear with $v$ or $u$, correspondingly, on the two
initial null segments. On the outgoing segment we take
$r_{,v}=1$. (Notice, that this implies that this $v$ is twice advanced
time at late times.) On the
ingoing segment, we take $r_{,u}={\rm const}\equiv r_{u0}$. (Notice that
$r_{u0}<0$.) The initial values of $r$ are thus uniquely determined by
the parameter $r_0\equiv r(u_i,v_i)$. We choose $u_{i}=0$ and
$v_{i}=r_{0}$, and thus we find: $r_v(v)=v$, and $r_u(u)=r_0+u
r_{u0}$. [Hereafter, we denote the initial values of the three fields
$r,\sigma,\Phi$ on the two segments of the characteristic hypersurface by
$r_v(v),\sigma_v(v),\Phi_v(v)$ and $r_u(u),\sigma_u(u),\Phi_u(u)$,
correspondingly.] Then, we can freely specify $\Phi_u(u)$ and $\Phi_v(v)$
(this choice represents a true physical degree of freedom). The initial
value of $\sigma$ is now determined from the constraint equations, namely
\begin{eqnarray}
\sigma_{u,u}=r_u(\Phi_{u,u})^{2}/(2r_{u0})\;\;,\;\;
\sigma_{v,v}=r_v(\Phi_{v,v})^{2}/2\;,
\label{initf}
\end{eqnarray}
together with the choice $\sigma(u_i,v_i)=-(1/2)\ln 2$. Thus, in the gauge
we use, we need to specify two functions of one variable [$\Phi_v(v)$ and
$\Phi_u(u)$] and two parameters ($r_0$ and $r_{u0}$) (in addition to
the charge $Q$) for the characteristic initial value problem.

{\it Determination of characteristic data}: We take the characteristic
data to satisfy $\Phi_v(v)=(A/\sqrt{-2r_{u0}})\ln (v/v_i)$ along $u=u_i$. 
Here, $A$ is a real constant, which is related to the amplitude of the
perturbation field. 
This choice for the scalar field is clearly of noncompact
support, as the field grows logarithmically in advanced time. (Notice,
that this implies that spacetime is not asymptotically flat.) 
In addition,
we require that $\Phi_{u,u}(u)=0$ on $v=v_i$, such that the field does not
propagate outside the event horizon on the ingoing segment of the
characteristic hypersurface. We also require that 
$\Phi$ is continuous at $(u_i,v_i)$. 

We note that is it unimportant what the field is along $v=v_i$ and on
$u=u_i$ for $v_i<v<v_f$ for any finite $v_f$: Any perturbation
field with compact support leads to power-law tails at late times
regardless of the specific shape of the initial data. It is only the
contributions of the characteristic initial data from late advanced times 
which are important. That is, one can approximate the characteristic
hypersurface by dividing it into two parts: a compact part which is
extended from $(u_i,v_i)$ to a point $(u_i,v_f)$ (with $v_f>v_i$), 
and a non-compact part which extends from $(u_i,v_f)$ forward, i.e., the
points
$(u_i,v>v_f)$. The specific form of the characteristic initial data on
$(u_i,v<v_f)$ is unimportant: it is only the contribution of the initial
data at $(u_i,v>v_f)$ which is important. (Similarly, also the initial
data along $v=v_i$ is unimportant.) Consequently, we can determine
arbitrary initial data at early times. 

The solution of the characteristic initial value problem then is
given by $\Phi_v(v)=(A/\sqrt{-2r_{u0}})\ln (v/v_i)$, $\Phi_u(u)=0$,
$\sigma_v(v)=-\ln [2(v/v_0)^{A^2/(2r_{u0})}]/2$, and
$\sigma_u(u)=-(1/2)\ln 2$.

\begin{figure}
\input epsf
\centerline{ \epsfxsize 8.5cm
\epsfbox{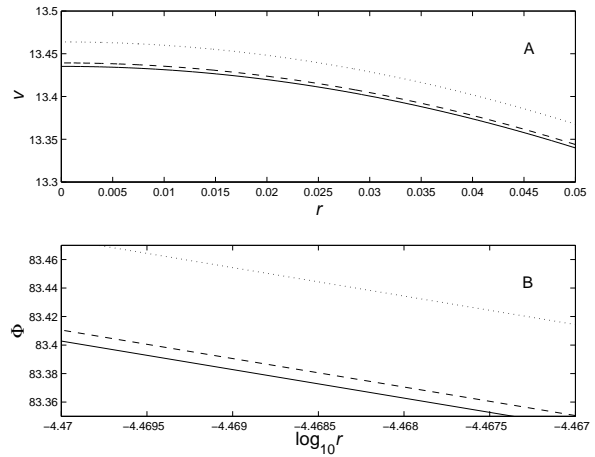}}
\caption{Behavior of $\Phi$ and $r$ along an outgoing ray (at fixed
$u$) for various values of $N$. Upper panel (A): $v$ as a function of $r$. 
Lower panel (B): $\Phi$  as a function of $r$. 
In both panels dotted lines correspond to $N=10$,
dashed lines to $N=20$, and solid lines to $N=40$, and the data are taken 
for $u=21.9$.}
\label{fig1}
\end{figure}

{\it Numerical simulations}: Our numerical code is a free evolution code
in $(1+1)$-D in double-null
coordinates with an adaptive mesh refinement \cite{burko-ori97}. We tested
the code and found that it is stable, and converges with second 
order. In the following we present results with the following choice of
parameters, unless stated otherwise: $M_{\rm initial}=1$, $Q=0.95$,
$r_0=5$ (these uniquely determine the value of $r_{u0}$), $A=0.3$, and
$N=10$. Here, $M_{\rm initial}$ is the initial mass of the black hole, and  
$N$ is defined as the number density of grid points on the characteristic
hypersurface in both $u$ and $v$ directions. We find similar qualitative
results also for other choices of the parameters.  
The stability and second-order convergence are demonstrated in
Fig.~\ref{fig1}, which displays $\Phi$ and $v$ as functions of $r$ along
an outgoing null ray deep inside the black hole for various values of the
grid parameter $N$. 

\begin{figure}
\input epsf
\centerline{ \epsfxsize 8.5cm
\epsfbox{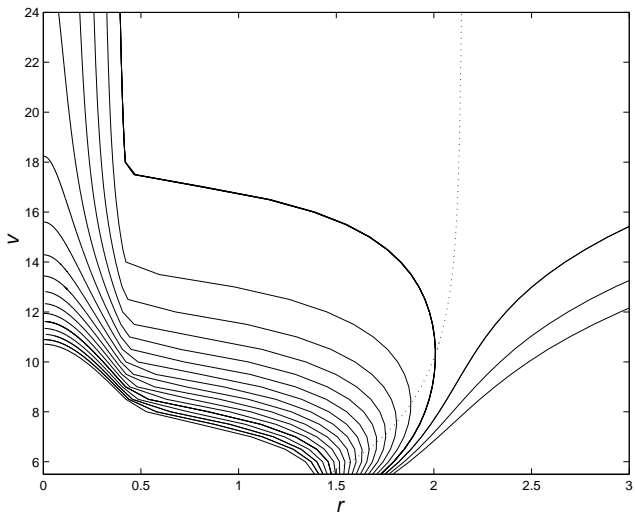}}
\caption{Outgoing null rays (with fixed values of $u$) in the
$rv$-plane. 
The solid lines correspond to different fixed values of $u$, 
and the
dotted line describes the apparent horizon, which approaches
$u=21.07185$ at late values of advanced time $v$. The values of $u$
for which rays are shown (from right to left) are from $u=20.8$ to
$22.7$ in equal increments of $\Delta u=0.1$.}
\label{fig2}
\end{figure}

Figure~\ref{fig2} displays equi-spaced (in $u$) outgoing null rays (with
constant values of $u$) in the $rv$-plane. The strong nonlinear
dynamics is demonstrated by the rapid increase in the apparent
horizon. All the rays which are not outside the event horizon and escape
to
infinity, either terminate at $r=0$ within a finite lapse of advanced time
$v$ (type I), or approach a finite limiting value of $r$ as $v\to\infty$  
of $r$ at large values of $v$ (type II). 

\begin{figure}
\input epsf   
\centerline{ \epsfxsize 8.5cm
\epsfbox{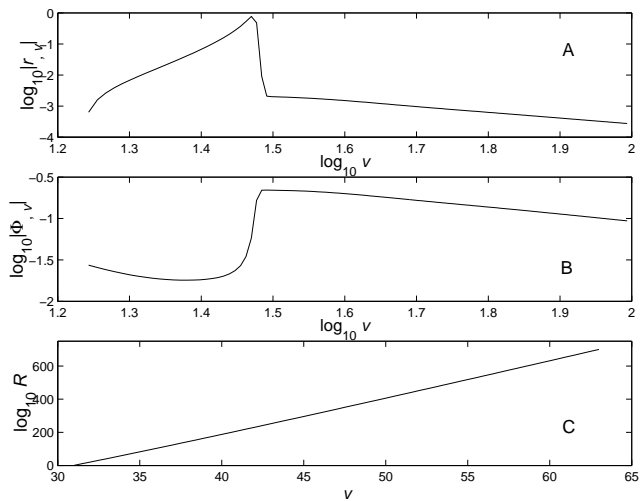}}
\caption{Behavior of the fields and the curvature of spacetime 
along a type-II outgoing null ray. Upper panel (A): $r_{,v}$ as a function
of $v$. Middle panel (B): $\Phi_{,v}$ as a function of $v$. Lower panel
(C): $R$ as a function of $v$.
The data are shown along $u=21.1$.}
\label{grads}
\end{figure}

{\it The null portion of the singularity}: Figure~\ref{grads}(A) shows
the behavior of $r_{,v}$ along a type-II outgoing ray. At late times
$r_{,v}\propto v^{-2}$. Figure~\ref{grads}(B) shows
the behavior of $\Phi_{,v}$ along the same outgoing ray. At late times
$\Phi_{,v}\propto v^{-1}$. This implies that along type-II rays $r$ indeed
approaches a non-zero finite value as $v\to\infty$, but $\Phi$ diverges 
logarithmically in the same limit. This behavior is in sharp contrast with
the behavior of $\Phi$ in the case of perturbations with compact support,
where $\Phi$, too, approaches a non-zero finite value. We next check the
detailed behavior of the fields along type-II rays. We find that type-II
rays terminate (in the infinite future as $v\to\infty$) at a curvature
singularity. This is demonstrated by Fig.~\ref{grads}(C), which shows the
exponential increase in  
$R\equiv (R_{\alpha\beta}R^{\alpha\beta})^{1/2}$ along the same
outgoing
ray, $R_{\alpha\beta}$ being the Ricci tensor. (Notice that here $R$ is
not the Ricci curvature scalar.) 
The finiteness of $r$ at
the singular hypersurface suggests that the non-central portion 
of the singularity is deformationally weak.

Note that the late-time behavior of $r_{,v}$ and $\Phi_{,v}$, as is clear
from Fig.~\ref{grads}, starts to dominate much earlier than in the case of
perturbations of compact support. In particular, no quasi-normal modes
(QNM) are visible. The reason for that is that the gradients of $\Phi$
decays here much slower than the tails in the case of perturbations of
compact support. Specifically, $\Phi_{,v}$ decays here according to an
inverse power-law with a smaller index than in the case of perturbations
with compact support. (In the latter case the index is $4$ for spherical
perturbations, whereas here we have an index of $1$.) Because the field is  
stronger, it starts dominating earlier, and overwhelms the
rapidly-decaying oscillations of the QNM. 

{\it The spacelike portion of the singularity}: Type-I rays terminate at a
spacelike singularity. 
The spacelike singularity inside a spherical charged black hole which is
perturbed by a scalar field was studied within a simplified homogeneous
model in Ref.~\cite{burko99}, where the pointwise behavior of the geometry
and the field was found. It was also shown in \cite{burko99} that 
approaching the spacetime singularity, the fully nonlinear and
inhomogeneous numerical solution (where the perturbation field had a
compact support on the characteristic hypersurface) was in full agreement
with the pointwise behavior. The study of the singularity in
Ref.~\cite{burko99} was local: no assumptions were made regarding the form
of the perturbation field on the characteristic hypersurface. We thus
expect that type-I rays terminate at a spacelike singularity whose
pointwise behavior is well
described by the singularity of Ref.~\cite{burko99}. Assuming
homogeneity, one finds approaching the spacelike singularity inside a
spherical charged black hole with a scalar field, that 
\begin{equation}
\Phi(r)=\sqrt{\beta +1}\ln r +O(r^{\beta})\, .
\label{phi}
\end{equation}
Here, $\beta >0 $ is a constant (which numerically can be found to depend
on $u$). Also, along an outgoing null ray one can show that, to the
leading order in $v_*-v$, 
\begin{equation}
r(u_0,v)=[d^2/(\beta +1)]^{1/4}(v_*-v)^{1/2}\, ,
\label{r}
\end{equation}
where $d$ is a gauge-dependent quatity (which depends on the scaling of
the temporal coordinate), and $v_*=v(r=0)$ along that null ray. 

\begin{figure}
\input epsf
\centerline{ \epsfxsize 8.5cm
\epsfbox{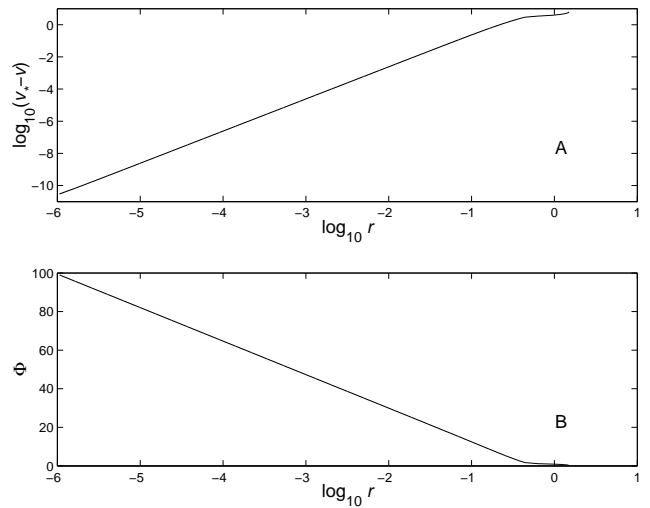}}
\caption{The scalar field $\Phi$ and the radial coordinate $r$ approaching
the singularity along a type-I null ray. Upper panel (A): $v_*-v$ as a
function of $r$. Lower panel (B): $\Phi$ as a function of $r$. The data
are shown along the outgoing null ray at $u=21.9$.}
\label{fig3}
\end{figure}

The agreement of our results with Eqs.~(\ref{phi}) and (\ref{r}) is
already apparent from Fig.~\ref{fig1}. Next, we check this agreement in
greater detail. In Fig.~\ref{fig3}(B) we show the
behavior of the scalar field $\Phi$ as a function of $r(v)$ along a type-I  
outgoing null ray. This logarithmic behavior is consistent with
Eq.~(\ref{phi}). Along all type-I rays we find the same
logarithmic divergence of $\Phi$, including along rays which initially
are outside or inside the apparent horizon. (The only difference between
different rays is that the slope of the graph, i.e., the value of the
parameter $\beta$, changes from one ray to another.)  
Figure \ref{fig3}(A) displays $v_*-v$ as a function of $r$ along
the same outgoing null ray. The asymptotic behavior approaching the
singularity agrees very nicely with Eq.~(\ref{r}).

We conclude that the pointwise behavior at the
singularity which we find
in our simulations is well described by the singularity described in 
Ref.~\cite{burko99}. Notice that this singularity is different from the
Schwarzschild singularity: The former has $\beta>0$, and the latter has
$\beta=-1$. This portion of the singularity then is scalar curvature,
spacelike, and deformationally strong. 

{\it Conclusions}: We studied the evolution of spacetime, and specifically
the formation of curvature singularities, for a very simplied toy model of
a spherical charged black hole, which is perturbed nonlinearly by a
self-gravitating, spherical scalar field, which has non-compact support on
the characteristic initial hypersurface. Although these perturbations are
stronger than those which result from an initial profile with compact
support (the gradient of the scalar field $\Phi$ decays at late times as
$v^{-1}$ in our case, and as $v^{-4}$ in the case of perturbations with
compact support) and consequently the evolution of spacetime and in
particular of spacetime curvature is indeed dominated by the non-compact 
perturbations rather than by the perturbations due to the collapse, we
find that a portion of the Cauchy horizon still survives as a non-central,
null singularity, rather than being utterly destroyed and replaced by a
central, spacelike singularity. The null generators of the Cauchy horizon
contract with retarded time $u$, and eventually arrive at $r=0$, where the
causal structure and the strength the singularity change: the central
singularity is spacelike and deformationally strong. This situation and
the global causal structure is therefore very similar to that of a black
hole perturbed by a perturbation field with compact support, despite the
different details of the dynamics. 

The reason why the Cauchy horizon survived the introduction of
non-compact perturbations as a null, non-central singularity is the
following: We chose the characteristic field to be such, that although it
is non-compact and does not belong to the same class of behavior on the
event horizon at late advanced time as the tails, its gradient
does. Specifically, along the event horizon $\Phi$ is logarithmic in
advanced time. This certainly does not belong to the class of the tails,
which decay as an inverse integral power of advanced time on the event
horizon. However, $\phi_{,v}$ decays like $v^{-1}$ along the event horizon
at late advanced time, such that it does belong to the same class as the
gradients of the tails. (Note, that no tails would ever produce $n=1$, as
$n$ is at least $3$ for all tails.) It is, in fact, $\phi_{,v}$ which is
the important quantity, as curvature depends on the gradients of $\Phi$,
rather than on $\Phi$ itself. This particular form of $\phi_{,v}$
implies that ${\cal R}$ can still be twice integrable approaching the
Cauchy horizon. 

We therefore conclude that by themselves, perturbations of non-compact
support, even when they dominate the dynamics, are not sufficient to
obliterate the null, non-central singularity at the Cauchy horizon. It
remains an open question, however, whether other classes of non-compact
perturbations behave similarly. The CBR perturbations,
which are a generic source of perturbations for realistic black holes, are
of particular interest. The energy influx of the CBR decays only
on very long time scales due to the expansion of the universe (in a
matter-dominated universe). (In a dark-energy dominated universe the
influx of CBR energy decays faster.) It is interesting to investigate how
the Cauchy horizon singularity is affected by such perturbation fields,
and also to investigate which families of perturbing fields may destoy
the null, non-central singularity.

I thank Karel Kucha\v{r} and Richard Price for discussions, and an
anonymous referee for useful comments. This research
was supported by the National Science Foundation through grant No.\
PHY-9734871.

\end{document}